\documentclass[twocolumn]{revtex4}
\usepackage{graphics}

\newcommand{\gae}{\lower 2pt \hbox{$\, \buildrel {\scriptstyle >}\over
{\scriptstyle
\sim}\,$}}
\newcommand{\lae}{\lower 2pt \hbox{$\, \buildrel {\scriptstyle <}\over
{\scriptstyle
\sim}\,$}}

\begin{document}

\title{Screening in anisotropic superfluids
and the superfluid density in underdoped cuprates}

\author{Matthew J. Case and Igor F. Herbut}

\affiliation{Department of Physics, Simon Fraser University, 
Burnaby, British Columbia, Canada V5A 1S6 }

\begin{abstract}
We examine the nature of the collective excitations
in a strongly anisotropic system of bosons 
interacting via Coulomb interaction. Such a system has often been used
in the past to
model the effects of quantum and classical phase fluctuations on the
superfluid density of underdoped cuprates.
Depending on the anisotropy and the effective strength of the
interaction we find
four different regimes for the temperature dependence of the
superfluid density. Coulomb interaction in underdoped cuprates 
is argued to be effectively short-ranged, and less then unity in
appropriately defined units.
\end{abstract}
\maketitle

 Temperature dependence of the superfluid density $\rho(T)$ is one of the
 key properties of a superfluid. This is particularly true in 
 high-temperature superconductors, where the linear low-temperature behaviour
 of $\rho(T)$ at optimal doping served as early evidence for the d-wave
 symmetry of the superconducting order parameter \cite{hardy}.
 Another intriguing aspect of $\rho(T)$ is its evolution with underdoping, 
 the effect of which may  be expected to increase the importance of
 fluctuations. Indeed, it has been argued that for a low value of
 $\rho(0)$ as is observed
 in the underdoped regime the further reduction of $\rho(T)$ with temperature
 should be primarily due to classical phase fluctuations \cite{kivelson},
 \cite{stroud}. This argument has been
 criticized for its neglect of Coulomb interaction, 
 which is expected to strongly suppress classical phase fluctuations
 below the plasmon energy and thus reinstate quasiparticles as the
 source of the linear temperature dependence of $\rho(T)$ at low $T$
 \cite{millis}.
 The competition between the quasiparticles and phase fluctuations for
 the form of 
 $\rho(T)$ becomes quite explicit in the so-called Ioffe-Larkin rule
 \cite{ioffe}
 \begin{equation}
 \label{ILrule}
 \rho^{-1}(T) = \rho_{qp}^{-1}(T)+\rho_{fluct}^{-1}(T), 
 \end{equation}
 where $\rho_{qp}(T)$ is the standard BCS quasiparticle contribution
 in a d-wave state,  and $\rho_{fluct}(T)$ is the
 (model-dependent) fluctuation component. This transparent result was first
 derived within the context of effective gauge theories of the t-J
 model \cite{lee}, but it may be expected to apply more generally to strongly
 fluctuating quasi-two-dimensional
 superconductors with $\rho(0) \ll \rho_{qp}(0)$ \cite{igor1}.

\begin{figure}[t]
{\centering\resizebox{80mm}{!}{\includegraphics{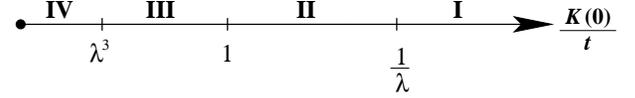}}\par}
\caption[]{ Four regimes for the temperature dependence of
the superfluid density $\rho(T)$ in
a layered bosonic system with weak Coulomb interactions: I)
quasi-two-dimensional (2D) regime with weak and effectively short-range
(screened) interaction for $K(0)/t\gg 1/\lambda$, II) quasi-2D regime
with weak long-range interaction for
$1\ll K(0)/t \ll 1/\lambda $ ,
III) three-dimensional (3D) regime with weak long-range
interaction for $\lambda^3 \ll K(0)/t \ll 1$,
and IV) 3D regime with strong long-range interaction for
$K(0)/t \ll \lambda^3$.
$K(0)=\hbar^2 \rho(0)/ m$, and $t$ is the inter-layer Josephson coupling.
$\lambda= (2 \pi e^2/(\epsilon d))/(\hbar^2 /(md^2))\ll 1$ is the dimensionless
strength of the Coulomb interaction, and $d$ is the inter-layer separation.}
\label{regimes}
\end{figure}

 Cuprates, however, are strongly anisotropic materials, with the
 anisotropy between the transport properties in the ab-plane and the
 c-axis increasing with underdoping.  It is well known that
 under these conditions the plasmon dispersion becomes very anisotropic as well,
 and the large plasmon energy gap gets replaced by a much
 lower one proportional to
 the interlayer coupling \cite{fertig}, \cite{paramekanti}.
 One may therefore expect Coulomb interaction in such highly anisotropic
 superconductors to become less efficient in gapping
 the phase mode. In this paper we study
 in greater detail the combined effect of large
 anisotropy and Coulomb interactions on $\rho(T)$.
 We model the fluctuation component in Eq.~\ref{ILrule} by a layered
 system of bosons interacting via Coulomb interactions
 and with the density proportional to doping. Such an effective theory
 arises naturally in several theories of underdoped cuprates
 \cite{lee}, \cite{igor1}, \cite{igor2},  \cite{zlatko}, and provides
 a rather general representation of a charged layered superfluid.
 We begin with the simple case of a two-layer system and demonstrate
 that, when the layers are Josephson-decoupled,
 there exists a linearly dispersing phase mode at low wavevectors.
 This is essentially a consequence of the perfect screening of interactions
 in one layer by fluctuations in the other. Generalizing to
 a system with an infinite number of layers, for a weak Coulomb interaction 
 we find that there are four discernible regimes for the temperature dependence of
 the superfluid density, controlled by the ratio between the boson
 density and the Josephson coupling (Fig. \ref{regimes}). An estimate
 of the relevant parameters places the underdoped YBa$_2$Cu$_3$O$_{6+x}$ (YBCO)
 with $T_c > 5K$ firmly in the regime I, where the Coulomb interaction acts 
 as an effective short-range interaction;
 the long-range nature of the Coulomb interaction in this regime is irrelevant
 except at extremely low temperatures. We also find that the value
 of the dimensionless interaction strength $\lambda$ in YBCO is
 $ \lambda < 1$, and argue that it is possible that $\lambda \ll 1$.

 Let us define the quantum mechanical action for a layered system of
 interacting bosons as $S=\int_0^\beta {\rm d}\tau \int {\rm d}^2\vec{x}{\cal L}$
 with
 \begin{eqnarray}
{\cal L}&=&\sum_{i=1}^N
\Phi_i^*(\vec{x},\tau)(\partial_\tau- \frac{\nabla^2}{2m} -\mu)
\Phi_i(\vec{x},\tau)\\
\nonumber
&&+ \frac{1}{2} \sum_{i,j} \int d^2 \vec{x}'
|\Phi_i(\vec{x},\tau)|^2 
V_{ij} (\vec{x}-\vec{x}') |\Phi_j(\vec{x}' ,\tau)|^2\\
\nonumber
&&-t\sum_i\sum_{j=i\pm1}\Phi_i^*(\vec{x},\tau)\Phi_j(\vec{x},\tau), 
\end{eqnarray}
where $N$ is the number of two-dimensional (2D) layers, $t$ is a weak Josephson
coupling between the layers,
$\beta=1/T$, and we
set $k_B=\hbar=1$ throughout, unless otherwise noted.  
The Coulomb interaction is
\begin{equation}
V_{ij}(\vec{x}-\vec{x}')
= e^2 /\epsilon \sqrt{|\vec{x}-\vec{x}'|^2 +
|i-j|^2 d^2} ),
\end{equation}
$d$ being the separation between the layers
and $\epsilon$ the static background dielectric constant.
We assume the presence of a neutralizing background of
density $\rho_0$ equal to the average areal density of bosons at the
chemical potential $\mu$.

The nature of the excitations in the above superfluid system is
most explicit in a system of only two layers, which differs already
from a single layer in an important way. Let us introduce
first the usual density-phase variables \cite{popov} as
$\Phi_i (\vec{x},\tau)= \sqrt{\rho_0 + \Pi_i (\vec{x},\tau)} 
e^{i \theta_i (\vec{x},\tau)}$ and expand the Lagrangian
to the quadratic order in $\Pi_i$ and $\theta_i$,
\begin{eqnarray}
{\cal L}&=&\sum_{i=1}^2\left( \frac{\rho_0}{2m} (\nabla\theta_i)^2 +
\frac{(\nabla \Pi_i)^2 }{8m \rho_0} + i (\rho_0 + \Pi_i)\dot{\theta_i}\right) 
\\ \nonumber
&&+ \frac12\sum_{i,j}\int{\rm d}^2\vec{x}'\Pi_i(\vec{x},\tau)
V_{ij}(\vec{x}-\vec{x}')\Pi_j(\vec{x}',\tau)\\ \nonumber
&&+ t\rho_0 (\theta_1 - \theta_2) ^2
+\frac{t}{4\rho_0} (\Pi_1 - \Pi_2)^2.
\end{eqnarray}
We assumed here the usual periodic boundary conditions in imaginary time:
$\Pi_i (\vec{x},\beta)= \Pi_i (\vec{x},0)$ and
$\theta_i (\vec{x},\beta)= \theta_i (\vec{x},0) + 2\pi n_i (\vec{x})$, with
$n_i (\vec{x})$ integer. By rotating the fields
as $\theta_{\pm} = (\theta_1 \pm \theta_2)/\sqrt{2}$,
$ \Pi _\pm = (\Pi_1 \pm \Pi_2)/\sqrt{2}$, we can decouple the above
Lagrangian as ${\cal L}={\cal L_+} + {\cal L_-}$, with
\begin{eqnarray}
{\cal L_+}&=&\frac{\rho_0}{2m} (\nabla\theta_+)^2 +
i (\sqrt{2} \rho_0+\Pi_+)\dot{\theta}_+
+ \frac{(\nabla\Pi_+)^2}{8m\rho_0}\\
\nonumber
&&+\frac12\int{\rm d}^2\vec{x}'\Pi_+(\vec{x},\tau)V_+(\vec{x}-\vec{x}')
\Pi_+(\vec{x}',\tau),\\
{\cal L_-}&=&\frac{\rho_0}{2m} (\nabla\theta_-)^2 + i\Pi_- \dot{\theta}_-
+ \frac{(\nabla\Pi_-)^2}{8m\rho_0}\\ \nonumber
&&+\frac12\int{\rm d}^2\vec{x}'\Pi_-(\vec{x},\tau)V_-(\vec{x}-\vec{x}')
\Pi_-(\vec{x}')\\ \nonumber
&&+2t\rho_0 \theta_- ^2
+ \frac{t}{2\rho_0} \Pi_- ^2,  
\end{eqnarray}
where $V_\pm (\vec{x}) = V_{11} (\vec{x}) \pm V_{12} (\vec{x}) $. 
The Gaussian integration over $\Pi_\pm$ yields two branches of
excitations with the energies
\begin{equation}
\omega_+ ^2 = \frac{k^2}{2m} ( 2\rho_0 V_+ (k) + \frac{k^2}{2m}),
\end{equation}
\begin{equation}
\omega_- ^2 = (\frac{ k^2}{2m}+2t)
( 2\rho_0 V_- (k) + \frac{k^2}{2m} + 2t), 
\end{equation}
where
\begin{equation}
V_\pm (k) = \frac{2\pi e^2}{\epsilon k}( 1\pm e^{-kd}). 
\end{equation}
The branch $\omega_+$ describes the usual two-dimensional plasmon,
$\omega_+\approx
\sqrt{4\pi e^2\rho_0k /\epsilon m}$ at low momenta.
The two layers oscillate {\it in phase} and, as a consequence, $\omega_+$ is
independent of the Josephson coupling.
The canonically conjugate variable to $\theta_+$ is
the sum of two densities, and therefore the Coulomb interaction affects 
the energy of this mode, as is usual in systems with long-range interactions.
In contrast, the conjugate variable to $\theta_-$ is the difference between
the two densities which can oscillate without any cost in Coulomb
energy. As a result, for $t=0$,
$\omega_- \approx (2\pi\rho_0 e^2 d /\epsilon m)^{1/2} k$ at low momenta, and
the dispersion of the lower branch is the same as if the system
had only a short-range interaction of strength $\sim 2\pi e^2 d/\epsilon$.
With $t=0$ the layers cannot exchange particles and the density
in each layer therefore may oscillate so as to perfectly screen the Coulomb
interaction in the other. The oscillations,  
however, are then {\it out of phase}, and consequently when $t\neq 0$
this mode becomes gapped, with  
$\omega_- \approx \sqrt{2\pi t \rho_0 e^2 d/\epsilon }$ at low momenta. We will
refer hereafter to this energy as the Josephson gap. 

\begin{figure}[t]
{\centering\resizebox{80mm}{!}{\includegraphics{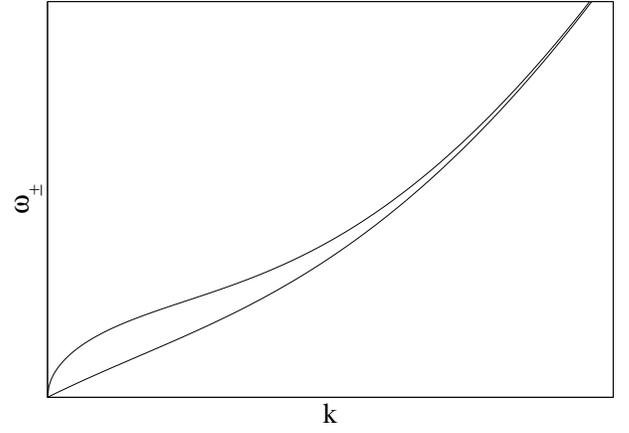}}\par}
\caption[]{The two branches $\omega_{\pm}$ of the excitation spectrum of the
two-layer system, with $t=0$ and weak Coulomb interaction.  The lowest mode crosses over from
linear behaviour at low $k$ to $k^2$ behaviour.  The plasmon starts out as
$\sqrt{k}$ before crossing to $k^2$.}
\label{2layers}
\end{figure}

The remarkable feature of the above result is that in a system
with negligible Josephson interaction between the layers, the Coulomb
interaction becomes effectively short-ranged as far as the low-energy
excitation spectrum is concerned. 
More precisely, when $t=0$, $\omega_- $ is linear and
deviates from $\omega_+$ significantly
for $k\ll 1/d$. For a large separation between the layers and 
for $1/d \ll k \ll
(8\pi m\rho_0 e^2/\epsilon)^{1/3}$,
$\omega_- \approx \omega_+ \sim \sqrt{k}$.
Finally, for $ (8\pi m\rho_0 e^2/\epsilon)^{1/3} \ll k$, $\omega_- \approx
\omega_+
\approx k^2 /2m$. If, on the other hand, the interaction is weak or if the
layers are
brought close together so that $1/d \gg (8\pi m\rho_0 e^2/\epsilon)^{1/3}$,
$\omega_- \sim k $ for
$k\ll \sqrt{8\pi m\rho_0 e^2 d/\epsilon }$, and
$\omega_- \approx k^2/2m$ otherwise, without the intermediate region
$\omega_ - \sim \sqrt{k}$.
In this regime $\omega_-$ becomes identical to the
phonon spectrum of the weakly interacting Bose gas. This is illustrated in Fig.
\ref{2layers}.  With a finite
Josephson coupling $\omega_-$ approaches the Josephson gap
for $k\ll \sqrt{4mt}$.  If the Josephson coupling were strong, of course,
$\omega_- \gg \omega_+$,
and the plasmon would resume its place as the low-energy mode of the system.

In a system with $N$ layers, for $t=0$ there are $N-1$
modes with linear dispersion and only a single plasmon. This is easily established 
by considering the interaction matrix $V_{ij}(k)$. When $k\rightarrow 0$,
$V_{ij}(k) = (2\pi e^2 /\epsilon k)(1+ {\cal O}(kd))$. So
in the limit $k \rightarrow 0$, 
\begin{equation}
\hat{V}(k) \rightarrow \frac{2\pi e^2}{\epsilon k} (1,1,...1)^T \otimes
(1,1,....1), 
\end{equation}
and the interaction matrix has one eigenvector with the eigenvalue $2\pi Ne^2/
\epsilon k$,
and $N-1$ degenerate eigenvectors with zero eigenvalue. The former eigenvector
is the total density which is
canonically conjugate to the sum of the phases and describes the plasmon.
The latter $N-1$ modes, being orthogonal to the plasmon, are electrically 
neutral and consequently cross from linear dispersion at low momenta
to the Josephson gap at $k=0$.

The existence of linearly dispersing modes below the usual plasmon
modifies the behavior of the superfluid density at low temperatures.
To be specific, we focus on the system
with infinitely many layers which is relevant in the context 
of high-temperature superconductivity. Imposing periodic
boundary conditions in the direction orthogonal to the layers the
excitation spectrum becomes
\begin{equation}
\omega^2 (k, k_z) = e(k,k_z) ( 2\rho_0 V(k,k_z) + e(k,k_z)),
\end{equation}
with $e(k,k_z) = ( k^2/2m) + t \sin^2 (k_z d/2)$, and \cite{fetter}
\begin{equation}
V(k,k_z) = \frac{2\pi e^2}{\epsilon k} \frac{ \sinh(kd)}{ \cosh(kd) -
\cos(k_z d) }.
\end{equation}
For $k_z =0$ one finds the usual three-dimensional plasmon at
$\omega ^2 (0,0) =  \omega_p ^2 =
4\pi e^2 \rho_0 / dm\epsilon$, while when $k_z \neq 0$ and $t=0$, for
$k \ll 1/d$,
$\omega(k, k_z) = \omega_p k/k_z $. The latter modes become gapped
when $t\neq 0$ and $\omega(k\rightarrow 0, k_z\neq 0) = \omega_p
\sqrt{t m d^2 /2 } $.

The temperature dependence of the areal in-plane superfluid density
in Landau's two-fluid model is given by 
\begin{eqnarray}
\label{dimful}
\rho(T)&=&\rho (0)\\ \nonumber
&&+ \frac{d}{2m} \int \frac{{\rm d}^2 \vec{k}}{(2\pi)^2}
\int_{-\pi/d}^{\pi/d} \frac{{\rm d}k_z}{2\pi}
k^2 \frac{\partial n_b (\omega(k,k_z)) }{\partial \omega(k,k_z)}, 
\end{eqnarray}
where $n_b(\omega)$ is the usual boson occupation number \cite{landau}.
We will find
it convenient to express the superfluid density in units of
energy by defining $K(T) = \hbar^2 \rho(T) /m$; we have also restored
the dimensionful Planck constant.  The rescaled
superfluid density $K(T)$ may then
be expressed entirely in terms of dimensionless quantities as
\begin{eqnarray}
\label{dimless}
\widetilde{K}(T)&=&\widetilde{K}(0)
- \frac{\widetilde{T}}{8\pi}\\ \nonumber
&&\times\int_0 ^\infty
y {\rm d}y
\int_0 ^1 {\rm d}z  \sinh^{-2}\Bigg\{\frac12
\Bigg{[}f(y,z)\Bigg(f(y,z)\\ \nonumber
&&+\frac{2\lambda \widetilde{K}(0) \sinh\sqrt{ 2 y \widetilde{T} } }
{ \widetilde{T}^{3/2} \sqrt{2y} ( \cosh \sqrt{2 y \widetilde{T} }
-\cos(\pi z)) }\Bigg{)}\Bigg{]}^{1/2}\Bigg\} , 
\end{eqnarray}
where $f(y,z)=y+(\tilde{t}/\widetilde{T})\sin^2(\pi z/2)$;
$\widetilde{X} = X/T_d$ are dimensionless, 
with $T_d= \hbar^2 /(md^2)$ as the
characteristic energy scale in the problem. The parameter
$\lambda = 2\pi e^2/(\epsilon d T_d)$
is the dimensionless measure of the Coulomb interaction's strength.

 Equation \ref{dimful} or \ref{dimless}, expected to be valid for
 $\lambda \ll 1$
 and not too close to the critical temperature, leads to four distinct
 regimes of temperature dependence of the superfluid density.
 Take $t/K(0) \ll 1$, as is
 relevant to the cuprates, and consider the function
 $K(T)$ as $K(0)$ is decreased at fixed $t$. We will 
 assume this to crudely correspond to underdoping
 a high-temperature superconductor, as we discuss shortly.

 \noindent
 I)  For $K(0) /t \gg 1$, the system is quasi-2D. Assuming
 $\lambda \ll 1$, to the zeroth order in $\lambda$
 the superfluid density in Eq. \ref{dimful} is easily seen to equal 
 the Bose condensate in the layered non-interacting system \cite{igor1},
 and
 \begin{equation}
 K(T) \approx K(0) - \frac{T}{2\pi} ((\ln \frac{T}{t}) + 1.386+ O(t/T)) , 
 \end{equation}
 over most of the temperature range. The deviations from Eq. 15
 are most significant
 below the crossover temperature $T_1 = \lambda K(0)$, where
 $\Delta K(T)=K(0)-K(T) \sim T^3$, and within the critical region of width $\sim
 \lambda T_c $ around $T_c$ with
 $T_c\approx 2\pi K(0)/\ln(2\pi K(0)/t)$. Besides $T_1$,
 there exists also a lower crossover
 temperature $T_1'$ of the order of the Josephson gap, where 
 $\Delta K(T)$ becomes exponentially suppressed. The latter temperature is
 \begin{equation}
 \frac{T_1 '}{T_c} = \frac{1}{2\pi} \sqrt{ \frac{\lambda t}{K(0)} }
 \ln\frac{2\pi K(0)}{t},
 \end{equation}
 whereas
 \begin{equation}
\frac{ T_1 }{T_c} = \frac{\lambda}{2\pi} \ln\frac{2\pi K(0)}{t}.
 \end{equation}
 The three characteristic temperature scales for variations of
 $K(T)$ will therefore satisfy the inequalities 
 \begin{equation}
\label{Tineq}
 T_1 ' \ll T_1 \ll T_c
 \end{equation}
 for
 \begin{equation}
 \frac{t}{K(0)} \ll \lambda \ll \frac{2\pi} {\ln(2\pi K(0)/t)}.
 \end{equation}
For such an interval for $\lambda$ to exist we obviously need
$t/K(0) \ll 2\pi/ \ln(2\pi K(0)/t)$, which is comfortably satisfied for
$K(0)/t >1$. When the inequality (\ref{Tineq}) is satisfied 
the long-range nature of the Coulomb interaction is
 relevant only at very low temperatures and the 
 interaction appears in $\rho(T)$  as being effectively
 short-ranged, and weak. Furthermore, as the ratio $K(0)/t$ is reduced the relative
 temperature range over which $K(T)$ behaves as a power law, 
 $T_1 /T_c$, decreases as well, albeit logarithmically slowly.
 One may interpret this decrease as that the interaction is being slowly
 renormalized towards zero with the reduction of the boson density
 \cite{fishoh}. The regime in which the inequalities
 (\ref{Tineq}) hold is marked as I in Fig. \ref{regimes}.

 \noindent
 II) If $K(0)$ is decreased further so that
 $K(0)/t \sim 1/\lambda$, one finds $T_1 ' \approx T_1\ll T_c$.
 The low-temperature behavior of the superfluid density
 corresponds to the long-range interaction. 
 This regime could be named the 2D, weakly interacting, long-range regime, and
 is labeled II in Fig. \ref{regimes}.

 \noindent
 III) For $K(0)/t\sim 1$, $K(T)$ to the zeroth order in $\lambda$, has the
 form of the fully three-dimensional Bose
 condensate over most of the temperature range, and 
 \begin{equation}
 K(T)\approx K(0) - 1.306 \frac{T^{3/2}}{\pi^{3/2} t^{1/2}}. 
 \end{equation}
 In this regime the exponential behavior of $\Delta K(T)$ sets in below
 \begin{equation}
 T_2 ' \approx \frac{\lambda^{1/2} }{ \pi }
 (\frac{t}{K(0)}) ^{1/6} T_c.
 \end{equation}
 The effective short-range behaviour
 $\Delta K(T) \sim T^4$, on the other hand, would appear below
 \begin{equation}
 T_2 = \lambda K(0) \approx \frac{ \lambda}{ \pi}
 (\frac{K(0)}{t})^{1/3} T_c.
 \end{equation}
 Note that $T_2/T_c$ now decreases as a power of $K(0)$, 
 which is again equivalent to the infrared renormalization group
 flow of the short-range coupling
 constant in a weakly interacting system of 3D bosons \cite{igor1}.
 However, since $T_2'> T_2$ for $K(0)/t <1/\lambda $, and $1/\lambda \gg 1$,
 by the time the system enters the 3D regime where $K(0)/t \sim 1$ the
 temperature dependence of $K(T)$ crosses over directly
 from exponential at low
 temperatures to $\sim T^{3/2}$ at higher temperatures. This is then the
 3D, still weakly-interacting, long-range regime, labeled III in Fig. \ref{regimes}.

 \noindent
 IV) Eventually, by reducing the density further one enters the
 regime where the Josephson gap becomes comparable to $T_c$. From Eq. 21
 this occurs when
 \begin{equation}
 \frac{K(0)}{t} \approx\frac{ \lambda^3 }{\pi^6 }. 
 \end{equation}
 This is the strongly-interacting regime (labeled IV in Fig. \ref{regimes})
 in which the superfluid density
 varies exponentially over the scale of $\sim T_c$. In this regime
 the system may be expected to eventually 
 suffer the phase transition into a Wigner crystal.

\begin{figure}[t]
{\centering\resizebox{80mm}{!}{\includegraphics{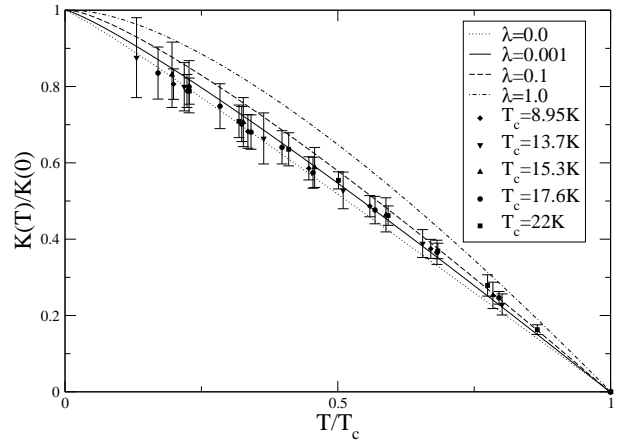}}\par}
\caption[]{ The scaled  \cite{remark}
superfluid density $K(T)/K(0)$ as a function of $T/T_c$ for various
values of the dimensionless interaction $\lambda$, and the
data on underdoped YBCO \cite{liang}.}
\label{lambdadep}
\end{figure}

   To determine the relevant regime for cuprates one needs an estimate
 of the dimensionless coupling constant $\lambda$.
 We find that $2\pi e^2/\epsilon d
 \approx 2500 {\rm K}$ assuming $\epsilon \approx 30$ \cite{kastner} and
 $d=12\hbox{\AA}$ in YBCO. The estimate of the temperature scale $T_d$ requires
 some assumptions on the relation between the superfluid
 density of the bosons and the measured superfluid density.
 For a finite interaction
 \begin{equation}
 K (0) < \frac{\hbar^2}{m} \rho_0, 
 \end{equation}
 where the right-hand side is the undepleted superfluid density of
 the non-interacting system. In the effective gauge theories of the
 t-J model \cite{lee},  or of the fluctuating d-wave superconductor
\cite{igor1}, density of bosons at low doping
 equals the density of holes, and thus $\rho_0 = x/a^2$, with $a$ the
 lattice constant in the ab-plane. So
 \begin{equation}
  T_d > \frac{(a/d)^2}{x} K (0).
  \end{equation}
  Since $(a/d)^2 \approx 0.1$ in YBCO it is convenient to take the doping
  $x=0.1$ as the reference point, so that $T_d > K (0)$ at this
  particular doping.
  Furthermore, the total superfluid density is related to the
  superfluid density of the bosons via the Ioffe-Larkin rule in Eq. 1,
  written more precisely as 
  \begin{equation}
  K_{tot}^{-1} (T) =  K_{qp}^{-1} (T) + K^{-1} (T)
  \end{equation}
  where
  $K_{qp}(T) = K_{qp}(0) - \alpha^2 (2 \ln(2)/\pi) (v_F/v_\Delta) T +O(T^2)$
  is the contribution from nodal quasiparticles, with $\alpha$ as the
  corresponding Fermi liquid parameter \cite{durst}.
  Expanding to the first order in $T$ one finds
  \begin{equation}
  K_{tot} (T) =  K_{tot}(0) - (Z\alpha)^2
  \frac{2 (\ln 2) v_F}{\pi v_\Delta} T,
  \end{equation}
  where the  $Z= K (0) / K_{qp}(0)$.
  Measurements of the superfluid density and thermal conductivity
  lead to an estimate $Z\alpha \approx 0.8$ at
  $x\approx 0.1$ \cite{sutherland}, and therefore
  \begin{equation}
  K (0) = \frac{K_{tot}(0)}{1-Z} \approx 5 K_{tot} (0),
  \end{equation}
  assuming conservatively that $\alpha = 1$. 
  Estimating $T_c \approx 65 {\rm K}$ at $x=0.1$ in YBCO, the interpolation
  of known results on the penetration depth yields
  $1/ \lambda^2 (0) \approx 50/\mu{\rm m}^2$ \cite{tami},
  which expressed in Kelvins \cite{case}
  leads to $K_{tot}(0) \approx 400 {\rm K}$. This leads to the estimate
  \begin{equation}
  T_d > 2000 {\rm K}.
  \end{equation}
  A similar value of the lower bound is
  obtained by repeating the exercise at optimal
  doping. The value of the dimensionless coupling $\lambda$ in YBCO is then
  \begin{equation}
  \lambda < 1.25.
  \end{equation}

There are at least two reasons to suspect that the value of the
interaction parameter $\lambda$ may lie 
significantly below our estimated upper bound in the last equation. First, 
for $\lambda\sim 1$ $K(0)$ would be well below the non-interacting
value of $\hbar^2\rho_0/m$, which would in turn yield a lower value
of $\lambda$.
Second, our estimate is evidently very sensitive to the value of $Z$ in
Eq. 28; assuming $\alpha<1$, for example,
 would bring $Z$ closer to unity and significantly
increase the energy scale $T_d$, and thus decrease the value of $\lambda$.
So the Eq. 30 should be understood as a very comfortable upper bound, with
$\lambda$ most likely laying well below it.

 The Josephson coupling $t$ may be related to
 the superfluid density along the $c$-axis $K^c (0)$ as
 \begin{equation}
 \frac{t}{2 T_d} = \frac{K ^c (0)}{K (0)}.
 \end{equation}
 Assuming that the measured superfluid density in very underdoped cuprates
 is dominated by the bosonic component, the above ratio is
 $10^{-4}$, and appears to become doping independent at low dopings
 \cite{hosseini}.  So, we estimate 
 \begin{equation}
 \frac{K(0)}{t} \approx 10^4
 \end{equation}
 in the underdoped regime.

 As the critical temperature changes from $T_c = 92 {\rm K}$ at optimal
 doping to $T_c = 9 {\rm K}$ in the extremely underdoped regime in YBCO, the $ab$-plane
 superfluid density changes by roughly two orders of magnitude
 \cite{liang}.
 Assuming a constant Josephson coupling in this range leads to
 the left-hand inequality in Eq. 19,
 $\lambda \gg t/K(0)$, being comfortably satisfied
 by our estimates, and the Coulomb interaction may be safely considered
 to be effectively short-ranged. Allowing the Josephson
 coupling to also decrease with underdoping, which may be closer to
 reality, only strengthens the above conclusion. 
 The right-hand inequality, however, would not be quite
 satisfied for $\lambda \approx 1$.
 As we argued, however, this is only an upper bound, and
 $\lambda$ may in fact be
 significantly smaller. To see the effect of the
 coupling strength on the form of the
 superfluid density, we plot in Fig. \ref{lambdadep} $K(T)/K(0)$ from Eq. 14
 for various values of $\lambda$ \cite{remark}, together with 
 the experimental points \cite{liang} on YBCO with
 $9 {\rm K} \leq T_c \leq 22 {\rm K}$.
 The best fit is achieved for $\lambda=10^{-3}$, although
 it is clear that any value $\lambda < 0.1$ would be almost equally good.
 This supports the recent proposal \cite{igor1} by one
 of us that the superfluid density in very underdoped cuprates is
 essentially the Bose condensate of the non-interacting layered bosonic
 system. The reader should also remember that whereas Eq. 14 yields the correct
 temperature dependence at low temperatures, it does not include the critical
 fluctuations within the critical region of width $\sim \lambda T_c$
 near $T_c$. These are known to modify the superfluid density into
 $K(T) \sim (T_c - T)^{\nu_{xy}}$ , with $\nu_{xy}\approx 0.67$,
 \cite{igor-zlatko}, and thus contribute to an additional
 rounding of the curve $K(T)$ from its non-interacting form.
 The absence of any such discernible critical region in the data in Fig. 3
 additionally supports our suggestion that $\lambda$ in YBCO
 may be rather small.

 If $\lambda\sim 1$, the system crosses
 from a 2D, short-range regime, for $K(0)/t \ll 1$, to a 3D long-range
 regime, for $K(0)/t \sim 1$, and the regimes II, III, and IV from
 Fig. 1, well separated for weak coupling, now overlap significantly.
 The regime I, however, even in this case remains wide and distinct.
 In fact, it is well known that in $^4$He, which is a strongly interacting
 Bose liquid, the variation of the superfluid density with temperature is
 well described by Landau's two-fluid model, except in the critical region.
 It thus seems likely that Eq. 14 would remain qualitatively
 correct over most of the temperature region even if the bosonic system
 is not quite weakly interacting, as long as it is reasonably far from
 solidification.

 In conclusion, we discussed the nature of the collective modes in a
 strongly anisotropic bosonic superfluid with Coulomb interaction
 between bosons. In particular, the influence of the anisotropic
 dispersion of these modes on the temperature dependence of the superfluid
 density was analyzed. Depending on relative values of  the 
 anisotropy and the interaction four different regimes can be discerned
 for a weak interaction. A crude estimate of relevant parameters for cuprates
 shows that the Coulomb interaction
 in underdoped YBCO may be considered to effectively be short-ranged
 and at least marginally weak, and the system to be quasi-two-dimensional.

We are grateful to S. Dodge, P. Lee, and S. Sondhi for useful discussions.
This work was supported by the NSERC of Canada.


\begin{thebibliography}{99}
\bibitem{remark} Strictly speaking the superfluid density in Eq. 14
obeys the simple scaling as in Fig. 3 only approximately. The residual
dependence on $K(0)$ (or $T_c$) at fixed $\lambda$ and $\tilde{t}$ 
is extremely weak, however.
Each curve in Fig. 3 is thus an accurate representation of
a family of curves for very different values of $K(0)$.
\end{thebibliography}
\end{document}